# Tonal Noise of Voluteless Centrifugal Fan Generated by Turbulence Stemming from Upstream Inlet Gap

M. Ottersten[a,b,*], H.-D. Yao[a], L. Davidson[a]

[a] Department of Mechanics and Maritime Science, Chalmers University of Technology, Gothenburg, Sweden
[b] Swegon Operation, Box 336, SE-401 25 Gothenburg, Sweden

___

Abstract

In this study, noise generation is investigated for a generic voluteless centrifugal HVAC fan at an off-design operation point where tonal noise increases. The simulations are performed by coupling IDDES with the FW-H acoustic analogy, and the experiments are conducted in a rig consisting of a plenum chamber and a reverberation room. In contrast to typical tonal noise sources induced by the fan blades, we find out that another predominant source is the turbulence stemming from the gap between the fan shroud and the inlet duct. The turbulence evolves along with the shroud and is swept downstream to interact with the top side of the blade leading edge. The interaction accounts for uneven surface pressure distribution on the blades. Moreover, the pressure is significantly unsteady near the shroud. The power spectral density (PSD) of the noise shows obvious tones at 273 Hz that is approximately equal to the difference of the blade passing frequency ($BPF_0$) and the fan rotation frequency. By coarsening the mesh resolution near the inlet gap and shroud, we artificially deactivate the gap turbulence in the numerical simulations and, consequently, detect that the tone at 273 Hz disappears completely. At this frequency, the PSD contours of surface pressure fluctuations are found potent at the inlet gap and the blade top side only if the gap turbulence is resolved. These findings indicate that the tonal noise source at 273 Hz is the interaction between the gap turbulence and blades. As the gap turbulence exists near the shroud wall upstream of the blades, the rotating wall introduces rotational momentum into the turbulence due to the wall friction. Hence the tonal frequency of the interaction is smaller than $BPF_0$ with a decrement of the fan rotation frequency. To the authors' knowledge, it is the first time that voluteless centrifugal fans are studied for the noise generation from the gap turbulence.

___

* Corresponding author, E-mail address: martin.ottersten@chalmers.se

# 1 Introduction

Sound quality is an important criterion for evaluating the comfort level of an indoor environment [1, 2]. To date, technologies for isolating the indoor environment from external noise (emitted by cars and airplanes, etc.) have achieved remarkable development [3]. On the other hand, the internal noise from heating, ventilation, and air conditioning (HVAC) systems is difficult to isolate. Low-speed centrifugal fans installed in the HVAC systems are known as dominant noise contributors. The noise can be reduced by placing silencers in ducts, but the silencers introduce additional skin friction and decrease the cross-sectional area of the flow. Besides, the silencers are mainly effective in absorbing broadband noise rather than tonal noise. They can be tuned to damp the tonal noise at specific frequencies, while the tuning is not valid for a wide range of frequencies [4].

The fan tonal noise generation is attributed to multiple causes. Pressure and density fluctuations on fan blades are identified as dominant sources in a large body of studies, e.g., by Ffowcs Williams and Hawkings [5]. Besides, centrifugal fans with volutes are found with tonal noise sources existing in blade downstream wakes as well as the gap between the blade and volute [6]. The tonal noise at the blade passing frequency (denoted by $BPF_0$) is generated from the flow due to the interaction between the blades and the volute [7]. The volutes, therefore, play an important role in the tonal noise generation. By contrast, in centrifugal fans without volutes (i.e., HVAC voluteless centrifugal fans), the tonal noise at $BPF_0$ is generated from a helical unsteady inlet vortex that interacts with the rotating blades near the backplate of the fan, as found in both simulations and experiments [8]. Another cause is inflow distortion, which leads to flow separation at the blade root near the backplate [9]. Based on this finding, flow obstructions were proposed to position upstream of the fan inlet [10]. The shape and location of obstructions were identified as the key parameters for noise reduction. As the present study focuses on the tonal noise, broadband noise sources (e.g., see Ref [11]) are not reviewed.

In an HVAC voluteless centrifugal fan, there is a gap (also termed clearance) between the rotating fan shroud and the stationary inlet duct. The pressure difference between the fan's inner and outer sides drives air to pass through the gap. As clarified by Hariharan and Govardhan [12], increasing the gap width worsens the blade aerodynamic performance. According to Lee [13], the gap gives rise to a local jet that aggravates flow separation near the shroud. This finding was further proven in later studies [14, 15]. The flow separation exists around the intersection of the shroud and blade trailing edge. It features intensive turbulence kinetic energy (TKE) [15]. As demonstrated in experiments [16, 17], the curved shape of the shroud also accounts for the flow separation. It is worth noting that the shroud skin friction leads to extra rotational momentum in the flow medium near the shroud walls. This effect is different from conventional blade vortex interaction (e.g., [18]), where the medium is quiescent.

The present article focuses on investigating the noise generated from the gap turbulent flow. To the authors' knowledge, this phenomenon in voluteless centrifugal fans is studied for the first time. Apart from experiments in a rig consisting of a plenum chamber and a reverberation room, the numerical simulations in the current study are carried out using a hybrid method of the improved delayed detached eddy simulation (IDDES) [19] and the Ffowcs Williams and Hawkings (FW-H) acoustic analogy [5], which will be elaborated on in Section 3. The IDDES is used in the flow simulation, and the FW-H for the noise prediction. In some previous studies [20,21], the unsteady Reynolds-averaged Navier-Stokes (URANS) equations were adopted for the flow simulation. It was found that the aerodynamic forces obtained using the URANS are in good agreement with experimental data. This method coupled with the acoustic analogy well predicts principal tonal noise but not the broadband noise. The reason is that the URANS cannot resolve transient small-scale flow fluctuations that are also noise sources, although large-scale unsteady contents are captured. The IDDES, by contrast, simulates most of fluctuations despite the modeling of sub-grid scale quantities. Hence, coupling the IDDES with the FW-H acoustic analogy enables the accurate prediction for both tonal and broadband noise, as demonstrated in previous studies on vehicle cooling fans [22,23] and a classical axial fan [24].

The main goal of the current study is to gain knowledge about the effect of the fan inlet gap on the fan aerodynamics and consequent noise generation, especially the tonal noise, under a specific off-design point where the tonal noise is found increasing in experiments. The reasons for the noise increase will be addressed by artificially controlling the gap turbulent flow in the numerical simulations. The artificial controlling is accomplished using a mesh-resolution coarsened treatment. The correlation between the noise sources and the noise will be quantified based on spectral analysis.

## 2  Application case

### 2.1  *Geometry and computational setup*

The voluteless centrifugal fan and its inlet duct investigated in this study are illustrated in Fig. 1a. The inlet duct has a trumpet shape with a variable cross-sectional area, which is narrowed to the minimum value near the fan inlet. The minimum diameter of the inlet duct is slightly smaller than the fan inlet diameter to form the clearance (i.e., the gap) between the stationary and rotating parts. As shown in Fig. 1b, the fan and inlet duct are placed within a downstream duct, and the inlet duct is connected to an upstream duct. This simple geometry layout is designed for the numerical simulations. The upstream and downstream ducts are utilized to replace the original experimental setup, which is presented in the next section. This simplification lessens the geometry complexity but reserves the principal flow and noise characteristics. Table 1 lists the geometry parameters signified in Fig. 1. There are 7 blades in the fan.

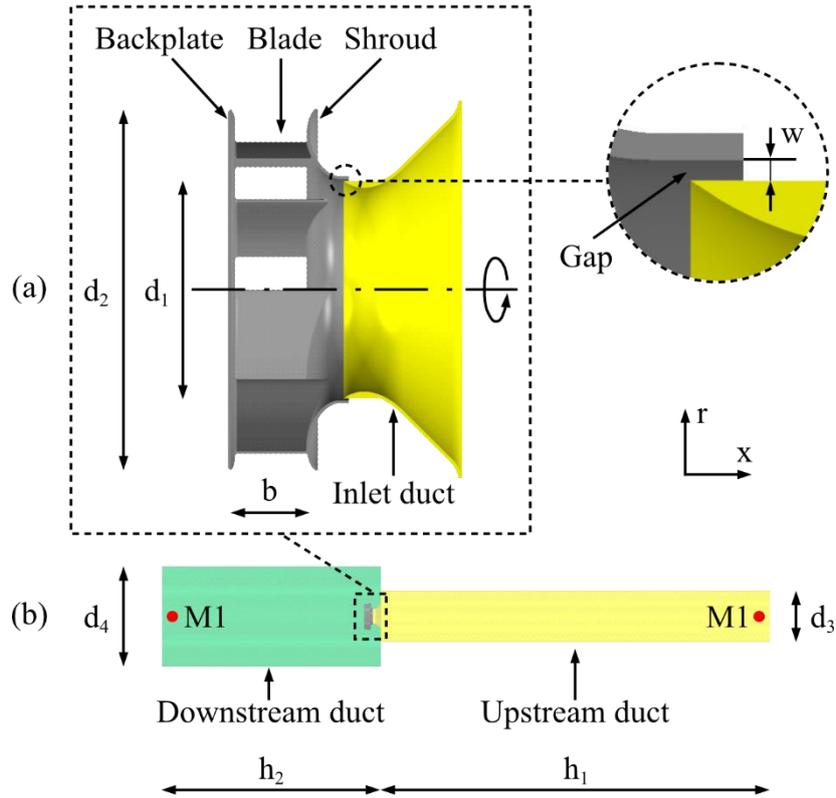

Fig. 1. a) The fan configuration and inlet duct. b) The simple geometry layout for the numerical simulations. Here M1 and M2 mark out two microphone positions. The rotation axis of the fan is the x-axis.

Table 1. The fan parameters.

| $d_1$ | $d_2$ | $d_3$ | $d_4$ | $b$ | $h_1$ | $h_2$ | $w$ |
|---|---|---|---|---|---|---|---|
| 0.165 | 0.268 | 0.6 | 1.1 | 0.053 | 4.0 | 2.3 | 0.0015 |

## 2.2 Experimental setup

The experimental rig for measuring the fan aerodynamics and aeroacoustics, which was validated in a previous study [15], is shown in Fig. 2. The rig consists of a reverberation room and a pressurized plenum chamber. The walls of the plenum chamber are treated with sound-absorbing materials. The fan is supported with a strut within the plenum chamber, and its trumpet-shaped inlet duct is connected to the reverberation room. The supporting strut is composed of slender rods that are positioned far away from the fan outlet, to eliminate the strut interference from the fan outflow. An electric motor is used to drive the fan. The motor rotating frequency is 46.7 Hz, which is much smaller than the blade passing frequency ($BPF_0$) and the harmonic frequencies of interest. The motor noise is measured without the fan at the rotating frequency so that the motor introduced noise is subtracted from the fan noise measurements.

By comparing Fig. 1b to Fig. 2, the reverberation room and the plenum chamber in the experimental rig are replaced with the upstream and downstream ducts in the computational setup, respectively. The replacement has been validated in the authors' previous study on the aerodynamics of the same fan using URANS [15]. It was found that the major aerodynamic characteristics and tonal noise frequencies measured in the experiments are reproduced by the simulations.

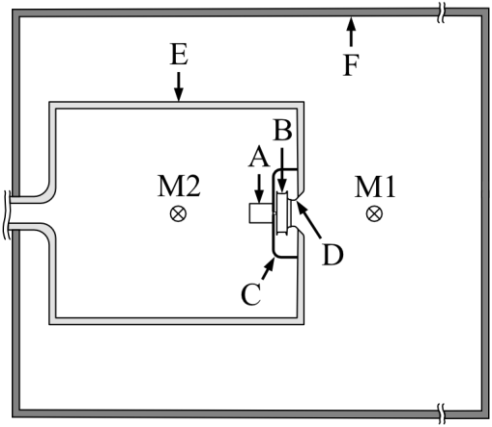

Fig. 2. The experimental rig: A – electric motor, B – fan, C – supporting strut, D – inlet duct, E – plenum chamber, F – reverberation room. M1 and M2 are microphones, which are identical to those in Fig. 1b.

The experimental parameters are given in Table 2. The microphones are PCB Piezotronics 130A20 40791. The software LabVIEW [25] is used to sample signals. The von Hann window is used for the fast Fourier transformation (FFT) in the spectral analysis. The sampling frequency is 25600 Hz, and the frequency resolution is 0.284 Hz.

Table 2. Experimental parameters.

| Density | 1.170 kgm$^{-3}$ |
|---|---|
| Temperature | 21.6 °C |

| Relative humidity | 35 % |
| Pressure in the reverberation room | 99360 Pa |

*2.3 Operation point*

The fan rotation speed is 2800 rpm (revolutions per minute). Given that the fan has 7 blades, the blade passing frequency $BPF_0$ is 326.7 Hz, and the first harmonic frequency 653.4 Hz.

The fan characteristic curve is shown in Fig. 3a. Five operation points are measured. The maximum efficiency point is found between Points 2 and 3, where the pressure rise is 410.83 Pa and the mass flow rate is 0.36 kg·s$^{-1}$. The power spectral density (PSD) of the tonal noise at $BPF_0$ is shown in Fig. 3b. The noise increases at the off-design condition Point 4, where the pressure rise is 269.65 Pa and the mass flow rate is 0.467 kg·s$^{-1}$. It was reported for voluteless centrifugal fans in [8] that the tone at $BPF_0$ increases with respect to the mass flow rate. Hence, Point 4 is specifically studied to understand the causes of the noise increase.

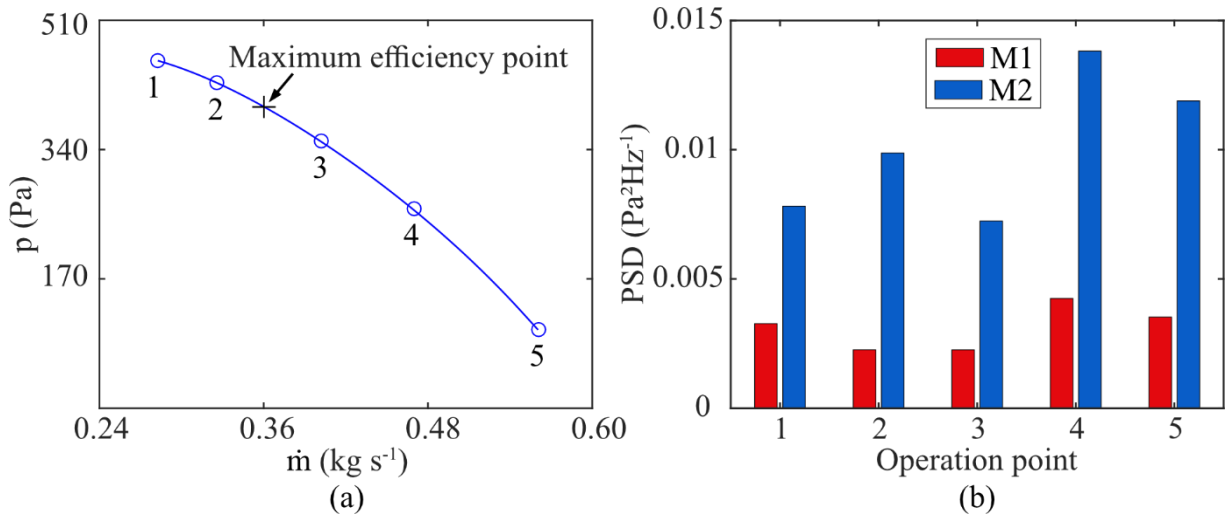

Fig. 3. a) The fan characteristic curve at the fan rotation speed of 2800 rpm. The operation points labeled with circles are further measured for the tonal noise. b) The PSD of the tonal noise at the blade passing frequency $BPF_0$ for the upstream microphone (M1) and the downstream microphone (M2).

## 3  Numerical methodology

*3.1 CFD method*

The air is considered as an ideal gas. The flow is compressible. A finite volume method is utilized to discretize the continuity, momentum, and energy equations. The method employs a segregated flow solver accomplished with the Semi-Implicit Method for Pressure-Linked Equations (SIMPLE) algorithm.

The convection flux on a cell face is discretized utilizing a hybrid second-order upwind and bounded scheme. The diffusion fluxes on both internal and boundary cell faces are discretized with a second-order scheme. The second-order hybrid Gauss-LSQ method is used in gradient computation. A second-order implicit method is taken to discretize the time derivative. The time marching procedure adopts inner iterations at every preconditioned pseudo-time step. All simulations are performed using the commercial software STAR-CCM+ [26]. This setup was validated in the previous study [27], where turbulence-induced acoustic waves transmitting through a cabin window were simulated.

The turbulence is simulated using the IDDES that is combined with the k-ω SST turbulence model. This setup has been tested in several studies on rotating machinery [22, 23]. The coefficients of the IDDES model adopts the default values in the software STAR-CCM+, i.e. $C_{DES,k-\omega} = 0.78, C_{DES,k-\varepsilon} = 0.61, C_{dt} = 20, C_l = 5,$ and $C_t = 1.87$. The notation of the coefficients is the same as in the software user guide [26]. The wall-normal sizes of the first layer cells near all walls fulfill $\Delta y^+ < 1$.

*3.2  FW-H acoustic analogy*

A hybrid approach is adopted to predict the noise generated from the flow. In this approach, the IDDES is coupled with Formulation 1A of the Ffowcs Williams and Hawkings (FW-H) acoustic analogy [28]. The ambient air density is set to 1.225 kgm$^{-3}$, and the speed of sound 340 ms$^{-1}$.

According to Neise [29], the fan noise generation at low Mach numbers is dominated by dipole noise sources that are derived based on the FW-H acoustic analogy. The same observation has also been reported for the current voluteless centrifugal fan using URANS coupled with Formulation 1A [15,20]. Hence, the noise prediction in this study considers only an impermeable integral surface for Formulation 1A. The integral surface consists of the fan blades, shroud, and backplate (see Fig. 1a), while the upstream and downstream ducts, as well as the fan inlet duct, are neglected. This treatment disregards the acoustic reflection from the duct walls to resemble the conditions in the experimental rig. The plenum chamber walls are installed with sound-absorbing materials, and the plenum chamber and reverberation room have large volumes for accommodating the sound wave propagation. Thus, there is a limited acoustic reflection from these room walls [20]. As the flow in the fan inlet duct does not contain fluctuations, noise sources on this surface are negligible. Nevertheless, an interesting future study is to investigate the duct effect on the noise propagation.

*3.3 Numerical settings*

The entire computational domain is divided into stationary and rotating parts. The parts contained within the ducts are stationary, whereas the part inside the fan is rotary. The meshes of the stationary and rotary meshes are not conformable at the interfaces between them.

The under-relaxation factors for the velocity and pressure in the segregated flow solver are set to 0.7 and 0.4, respectively. The under-relaxation factor for the turbulence equations is 0.7.

The mass-flow boundary condition is set at the inlet, with a uniform velocity distribution. The modeled turbulence intensity is set to $I = 4\%$ according to $I = 0.16(R_e)^{-1/8}$ [26]. Here $R_e$ is computed based on the inlet diameter and the streamwise velocity at the inlet. The modeled turbulence length scale is set to $\ell = 0.05$ m based on $\ell = 0.7d_3$ where $d_3$ is the upstream duct diameter. The pressure-outlet boundary condition is set at the outlet with the static pressure of 101325 Pa, which is the reference pressure ($p_{ref}$) in the ambient air. The no-slip boundary condition is specified on all walls.

The influence of convective Courant numbers on the simulation accuracy is investigated by setting two different time step intervals. The small time interval is $\Delta t_A = 2.0 \times 10^{-6}$ s, leading to the maximum convective Courant number below one and 10800 time steps per revolution of the fan. The large time interval is $\Delta t_B = 10\Delta t_A$ with the maximum convective Courant number of approximately 10. This value still fulfills the numerical stability required for the implicit time-marching method. For both time intervals, the maximum number of inner iterations per time step is set to 20.

The sampling period of the noise is 0.3 s for all cases, corresponding to 14 fan revolutions. The sampling time interval is $50\Delta t_A$. The PSD is calculated using the von Hann window for 3000 samples per signal section, which leads to a frequency resolution of around 3 Hz, and no overlapping between the signal sections.

Four cut planes (Planes 1-4) across the fan are specified to observe the flow quantities in the subsequent analysis. In addition, four monitoring points (P1, P2, P3, and P4) are located at a selected blade leading edge. The cut planes and points are shown in Fig. 4.

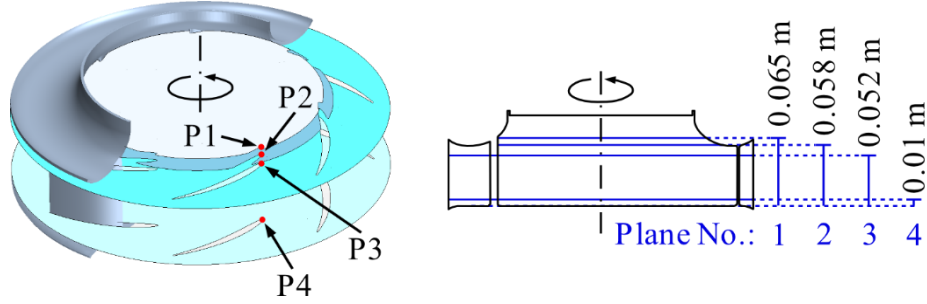

Fig. 4. The cut planes and points (P1-P4) for observing the flow variables in the subsequent analysis. The distances of the cut planes to the fan backplate are labeled.

## 4 Computational meshes

### 4.1 Local mesh refinement

We adopted a polyhedral mesh generation method to produce prism layers near the walls and polyhedral cells in the rest of the computation domain. The use of polyhedral cells for turbomachines has been demonstrated in [20, 30]. The growth rate is set to 1.05, as suggested in [31].

The mesh quality is evaluated using coarse and refined meshes (see Fig. 5). Local refinement is made inside the fan and inlet gap, where high TKE was found due to the gap turbulence [15]. The refined resolution enables the LES mode for the IDDES. The mesh parameters are listed in Table 3.

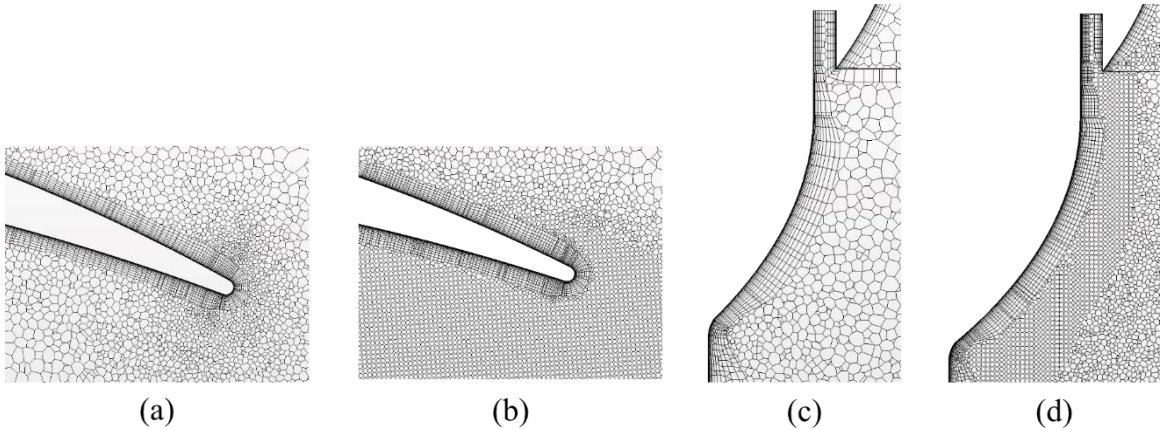

Fig. 5. Mesh cells near the blade trailing edge in a) the coarse mesh and b) the refined mesh; those near the inlet gap in c) the coarse mesh and d) the refined mesh.

Table 3. The mesh parameters.

|  | Coarse mesh | Refined mesh |
| --- | --- | --- |
| Total number of cells | $32 \times 10^6$ | $52 \times 10^6$ |
| Number of cells in the rotating zone | $22.9 \times 10^6$ | $41.9 \times 10^6$ |
| Maximum $\Delta y^+$ near blade walls | 0.93 | 0.73 |
| Cell growth ratio | 1.05 | 1.05 |

*4.2 Evaluation of mesh quality*

By setting different time steps of $\Delta t_A$ and $\Delta t_B$ for the coarse and refined meshes, four simulations are performed. The simulation matrix is shown in Table 4. Case 3 is simulated with the refined mesh and the small time interval, which lead to the global maximum convective number below one.

Table 4. The matrix of the simulated cases.

| | Mesh quality | Time interval |
|---|---|---|
| Case 1 | Coarse | $\Delta t_A = 2.0 \times 10^{-6}$ s |
| Case 2 | Coarse | $\Delta t_B = 10\Delta t_A$ |
| Case 3 | Refined | $\Delta t_A$ |
| Case 4 | Refined | $\Delta t_B$ |

The static pressure excluding the reference pressure ($p_{ref} = 101325$ Pa) is displayed in the axial symmetric line of the fan in Fig. 6. All cases show consistent pressure amplitudes upstream of the fan, while differences are seen downstream due to the mesh qualities. The influence of the time intervals is subtle given the same mesh is used.

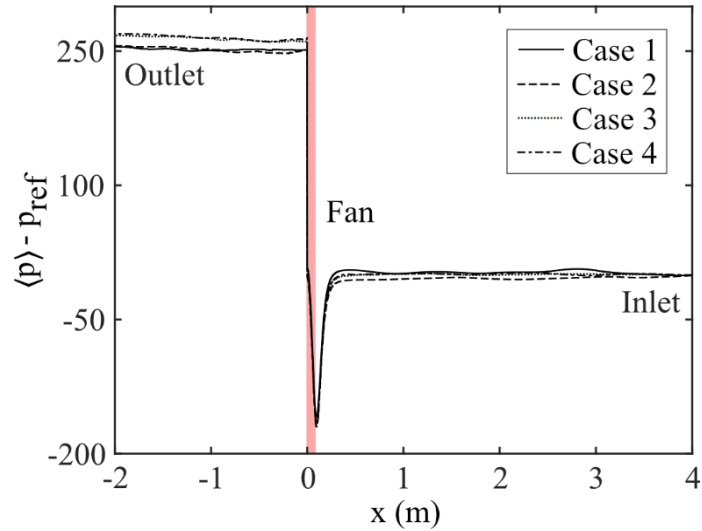

Fig. 6. The pressure in the axial axis of the fan across the computational domain. Here $x = -2$ corresponds to the location near the outlet and $x = 4$ near the inlet. The fan location is marked out with the red zone.

Important physical quantities describing the fan performance are the static pressure rise, which is between the fan inlet and outlet, and the fan torque. These two quantities from the four cases are listed in Table 5. The results of Cases 3 and 4 agree well with the experimental data.

Table 5. Fan performance data.

|  | Static pressure rise (Pa) | Torque (N·m) |
|---|---|---|
| Case 1 | 256.33 | 1.133 |
| Case 2 | 255.61 | 1.133 |
| Case 3 | 269.20 | 1.126 |
| Case 4 | 267.82 | 1.127 |
| Experiment | 269.65 | 1.125 |

# 5 Results and discussion

## 5.1 Inlet gap turbulence

The contours of vorticity magnitudes $\|\vec{\omega}\|$ near the fan inlet gap are shown in Fig. 7. In contrast to the coarse-mesh case (Case 2), the refined-mesh case (Case 4) resolves turbulent vortices that originate from the gap. The reason is that the fine mesh resolution turns on the LES mode of the IDDES [32]. The gap turbulence is mixed with the main flow as it is swept downstream. This phenomenon was also observed previously [13]. The turbulence is responsible for the noise generation [33]. Note that the instantaneous flow quantities analyzed throughout this paper are extracted at the same physical time.

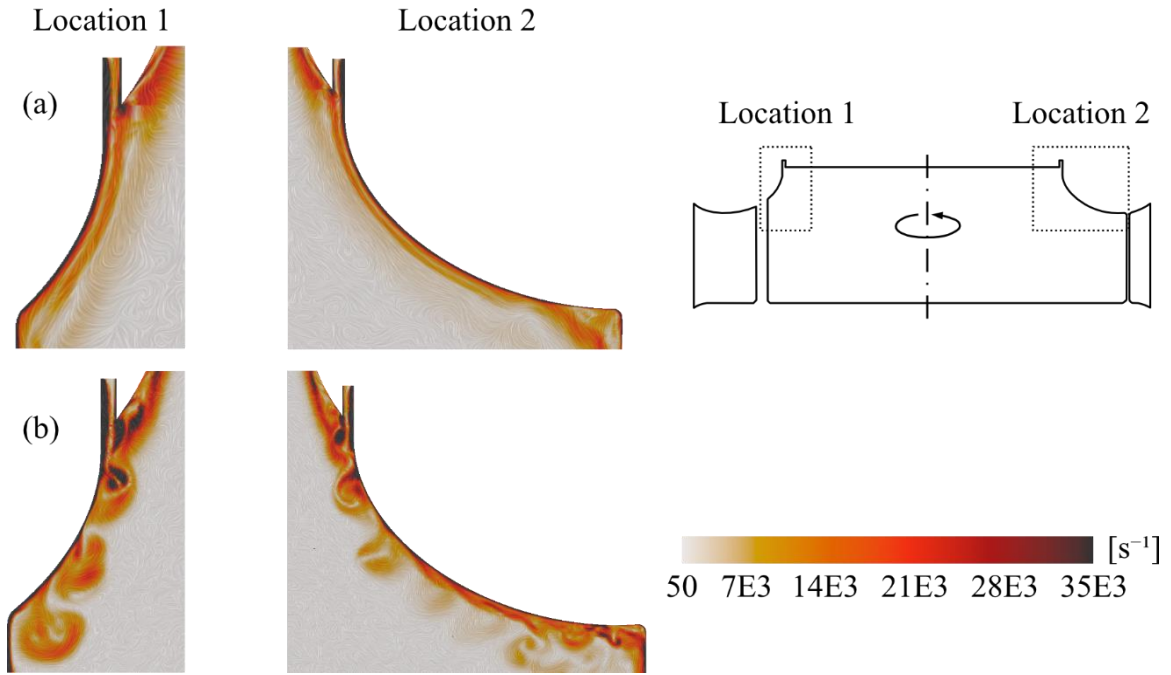

Fig. 7. Instantaneous vorticity magnitudes near the inlet gap. a) Case 2 with the coarse mesh; b) Case 4 with the refined mesh.

Furthermore, the vorticity magnitudes are shown in Plane 3 in Fig. 8. Plane 3 is positioned on the top side of the fan blades (see Fig. 4). More turbulent structures are resolved in Case 4 owing to the refined mesh quality. The turbulence is significant near the blade trailing edge on the suction side. The observation is consistent with the results obtained from the URANS [15,20], where large modeled TKE was found in the same regions, and the experiments [16,34].

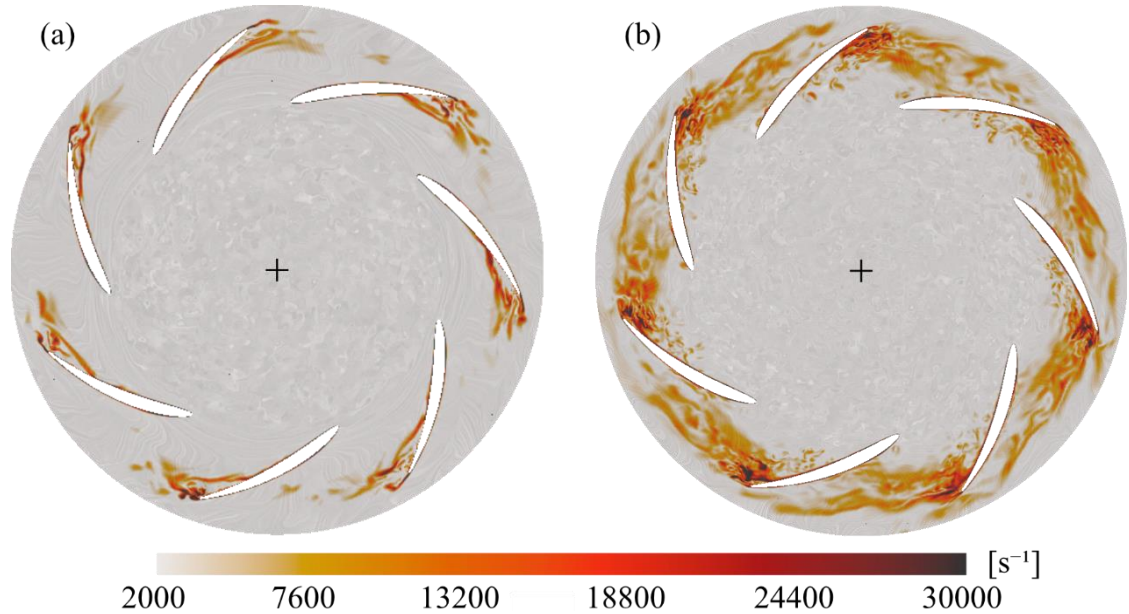

Fig. 8. Instantaneous vorticity magnitudes near the shroud are visualized in Plane 3: a) Case 2 without the inlet-gap turbulence and b) Case 4 with the inlet-gap turbulence.

Figure 9 displays the instantaneous surface pressure on the blade leading-edge side in Plane 2. Here $\eta_{chord}$ is the nondimensional distance to the leading edge that is calculated based on the blade chord length. The reference pressure, $p_{ref}$, is excluded from the static pressure shown in the figure. The pressure distribution among the blades are similar in Case 2, while noticeable discrepancies are observed in Case 4, for which the refined mesh is used. A common pressure peak is seen on the blade suction side at the position $\eta_{chord} = 0.018$. Case 4 exhibits that the peak varies in a wide range of approximately 200 Pa with the maximum value on Blade 1 and the minimum one on Blade 2. The pressure discrepancies among the blades are ascribed to the gap turbulence. Since the mesh is refined in the inlet gap and near the shroud wall, the turbulence from the inlet gap is well resolved. The turbulence develops along the shroud wall. It interacts with the blades at their leading edges. The interaction renders the uneven surface pressure distributions among the blades as well as the significant peak differences. The refined mesh is sufficient for resolving instantaneous turbulent fluctuations required for the noise prediction. As Case 4 with a larger time interval consumes less computational costs, it is selected for the following aeroacoustic analysis.

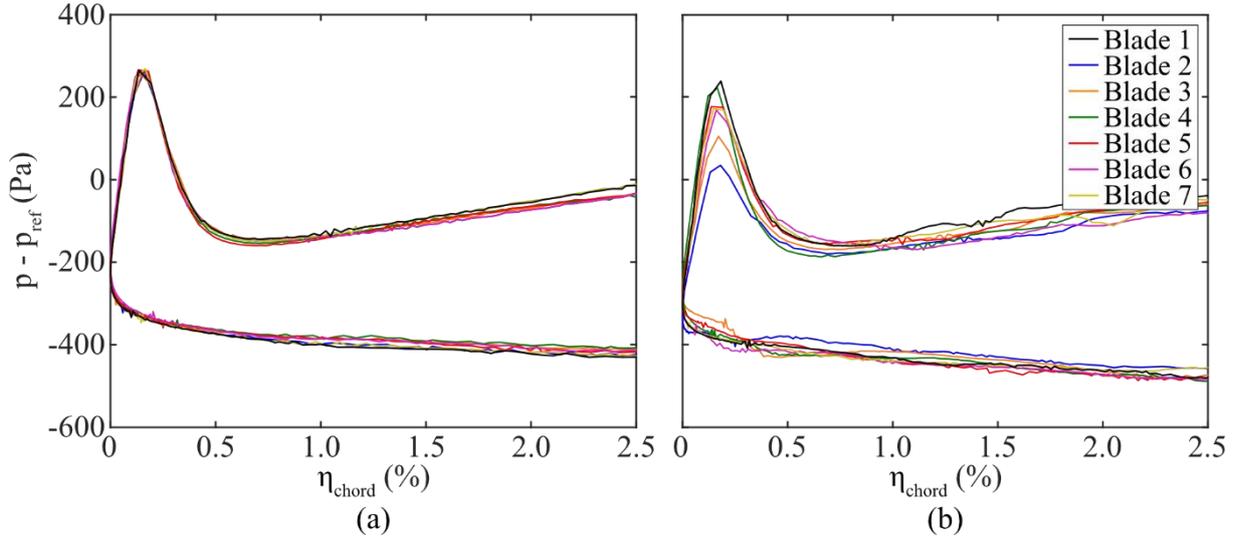

Fig. 9. The instantaneous surface pressure on the blades in the axial section, Plane 2, near the fan inlet: a) Case 2 with the coarse mesh, b) Case 4 with the refined mesh. Here $\eta_{chord} = 0$ corresponds to the blade leading edge.

*5.2 Interaction between inlet-gap turbulence and blades*

A snapshot of the vorticity magnitudes, $\|\vec{\omega}\|$, and the wall shear stress (WSS) magnitudes near the fan shroud is illustrated in Fig. 10. There are more isosurfaces of $\|\vec{\omega}\| = 2 \times 10^4 \text{ s}^{-1}$ near Blade 1, as compared with the neighboring Blade 2. The phenomenon of large vorticity magnitudes was also found in the previous study [35]. The isosurfaces imply intensive turbulent flow structures, which are developed from the inlet gap. These turbulent flow structures are swept past the blade leading edges. As a consequence, large WSS magnitudes are formed on the leading edge surfaces. As shown in Fig. 9b, the largest pressure peak exists on Blade 1 at the snapshot time, whereas the smallest pressure peak on Blade 2. The peak values are related to the intensive levels of the turbulence. The smallest pressure peak is found on Blade 2 since the most intensive vortices occur near this blade.

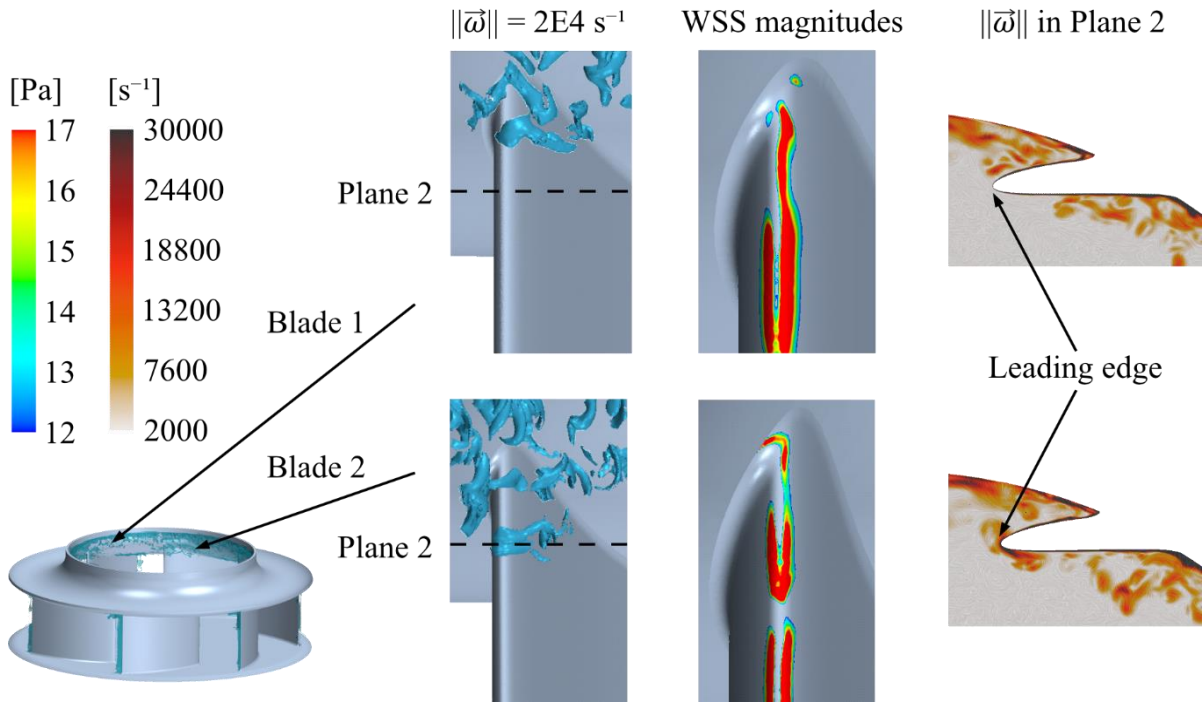

Fig. 10. A snapshot of the vorticity magnitudes, $\|\vec{\omega}\|$, and the WSS magnitudes. Note that the snapshot is taken at the same time as Fig. 9.

The instantaneous surface pressure on the blade leading edges at different fan axial positions are displayed in Fig. 11. At the position nearest the inlet gap, a remarkable pressure peak is seen at Blade 5, but the peaks at the other blades are similar. The blades have different pressure distributions on the suction sides near the shroud. However, the differences are minor at the positions that are far from the shroud. As the distance to the inlet gap and the shroud increases, the pressure differences among the blades decay. The reason is that the turbulence from the inlet primarily develops along with the shroud. The influence of the turbulence on the blades is, therefore, effective mainly near the shroud. Moreover, this physical behavior is explainable in terms of surface pressure contours (see Fig. 12), for which $p_{ref}$ is excluded. Blades 2 and 5 are chosen to compare. A region of large surface pressure is seen on Blade-5 leading edge near the shroud. But the surface pressure in the same region on Blade 4 is relatively small. Plane 1 crosses this region. Thus, the largest pressure peak in Plane 1 occurs on Blade 5, as also found in Fig. 11a.

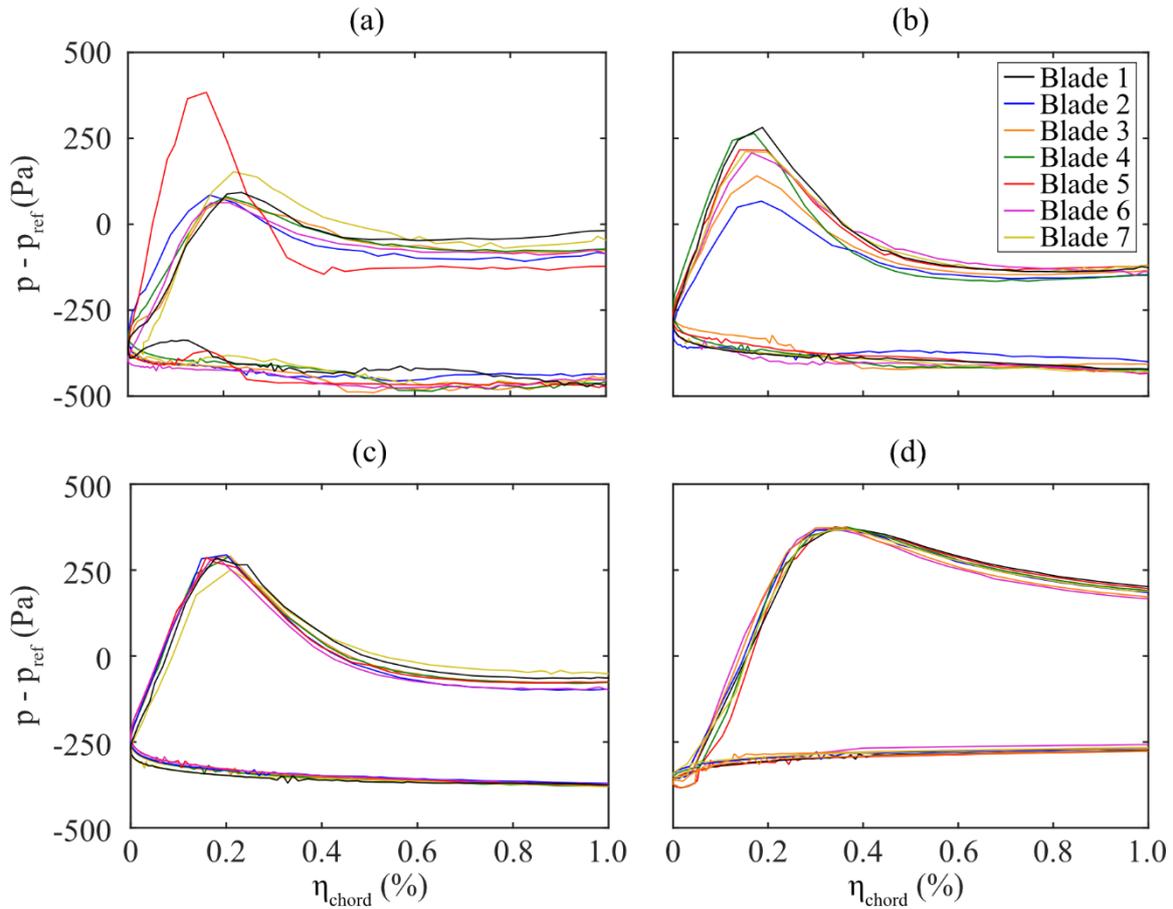

Fig. 11. The instantaneous surface pressure on the blade leading edges: a) Plane 1 closest to the inlet gap, b) Plane 2 intersecting with the fan shroud, c) Plane 3 without the intersection to the shroud, and d) Plane 4 near the fan backplate, which is furthest from the inlet gap (see Fig. 4).

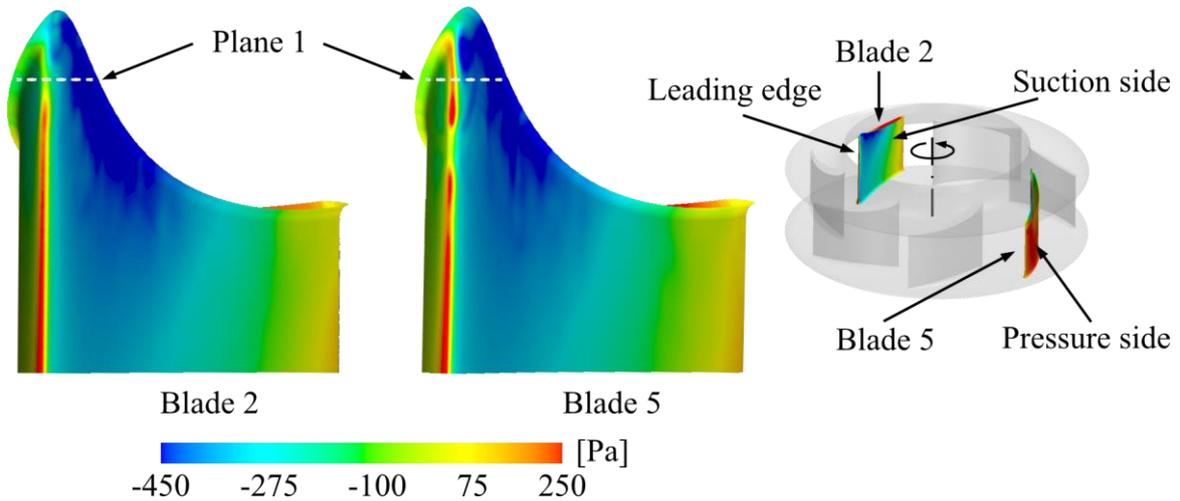

Fig. 12. The instantaneous surface pressure on Blades 2 and 5, viewed from the leading edge of the blade. The snapshot is taken at the same time as Fig. 11.

The time history of the surface pressure on Blade-2 leading edge at the monitoring points, Points 1-4, is displayed in Fig. 13. The first two points are located on the blade's top side near the shroud, and the others on the blade's bottom side (see Fig. 4). At Point 1, the pressure fluctuations with large amplitudes and high frequencies are observed. Fluctuations are also obvious at Point 2, but the fluctuating range is smaller. Moreover, a periodic low-frequency fluctuation in relation to the fan revolution is discerned, which was also found in [15]. By comparing the four monitoring points, high-frequency fluctuations decay rapidly with the increased distance to the shroud. The periodic low frequency becomes predominant at Points 3 and 4. The phenomenon is explainable based on the finding in Fig. 10. The gap turbulence gives rise to intensive surface pressure fluctuations. Since the gap turbulence fades on the blade bottom side, high-frequency fluctuations disappear. The periodic low frequency at Point 2 implies that although the maximum pressure peak is observed on Blade 5 rather than Blade 2 in Fig. 11a, the peak periodically moves among the blades. A detailed study of the periodic peak motion has been explored in [15].

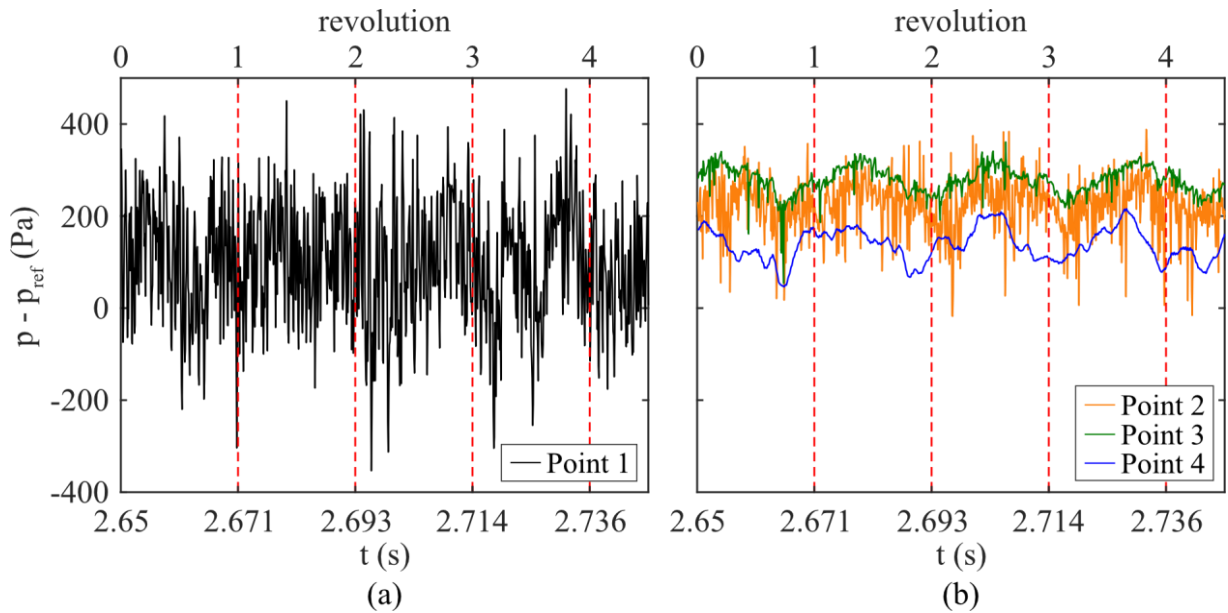

Fig. 13. The time history of the surface pressure on Blade 2 at the monitoring points along the blade leading edge at the intersection of a) Plane 1 (Point 1) and b) Plane 2 (Point 2), Plane 3 (Point 3), and Plane 4 (Point 4). The point locations are illustrated in Fig. 4. The Red dashed line indicates the fan revolution periods.

The surface pressure on the pressure and suction sides of Blades 3 and 6 is shown in Fig. 14. The pressure distributions of the two blades are overall similar. The only noticeable difference is that larger pressure exists on the pressure side of Blade-3 trailing edge. This effect is attributed to the gap turbulence that unevenly occurs among the blades. The same finding has been presented in the previous numerical simulations for centrifugal fans [15] and experiments for centrifugal compressors [16,34].

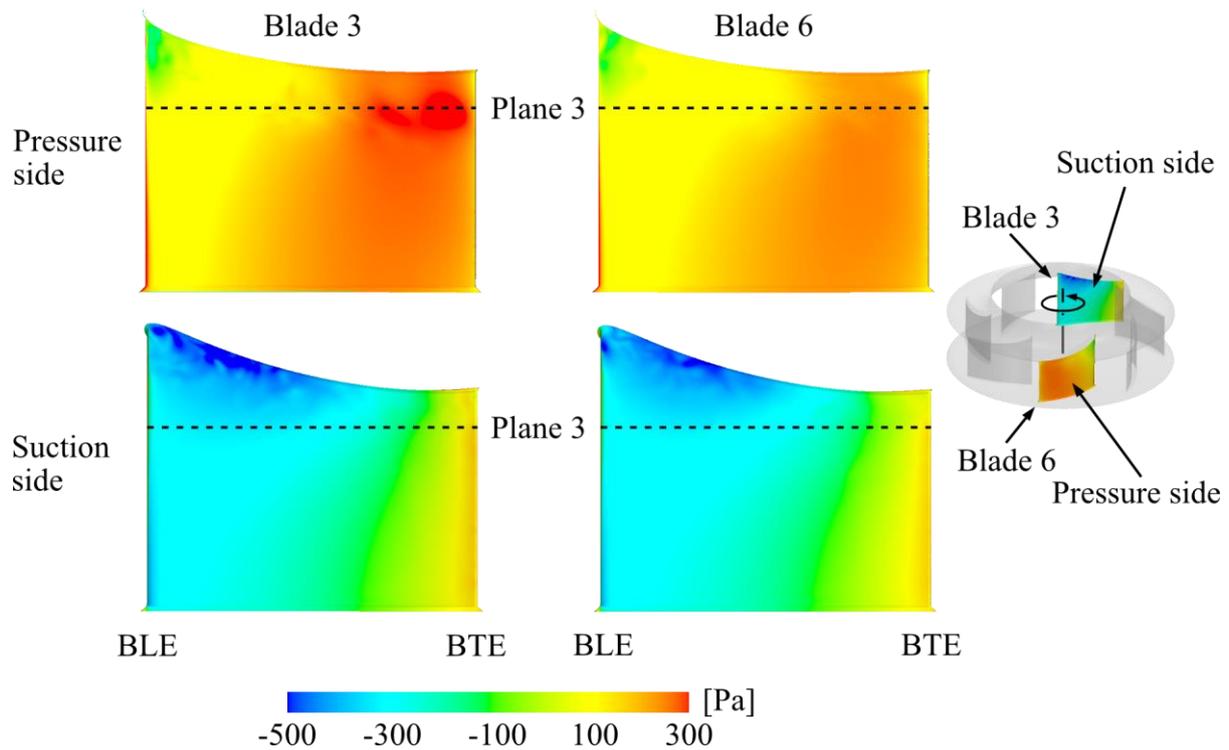

Fig. 14. A snapshot of the instantaneous surface pressure on Blade 3 and Blade 6. BTE is the blade trailing edge, and BLE is the blade leading edge.

The instantaneous surface pressure over the whole blade surfaces in Plane 3 near the shroud is displayed in Fig. 15. The pressure on the pressure side of the Blade 6 trailing edge is lower than the other blades. This result is explained based on the pressure contours in Fig. 14. Except for Blade 6, the other blades (e.g., Blade 3 shown in the figure) are subjected to low pressure on the blade pressure sides near the trailing edges. The low pressure region is caused by flow separation according to the previous studies [15, 20]. .

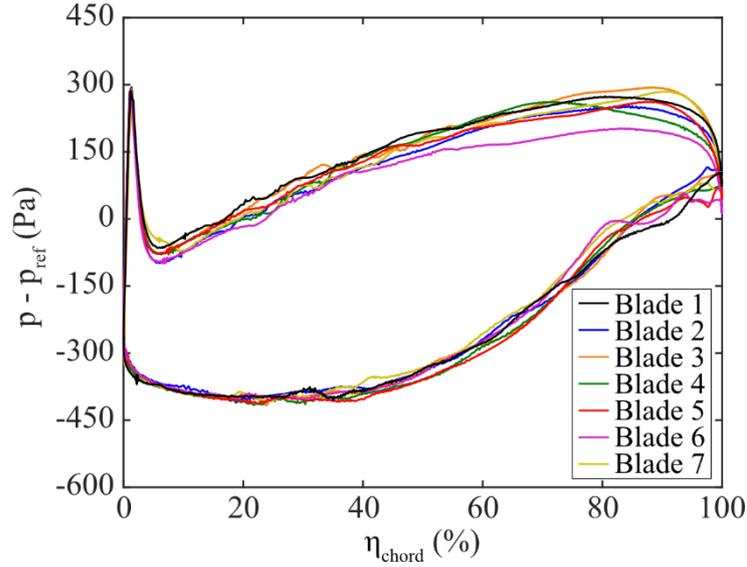

Fig. 15. The instantaneous surface pressure on the blades in Plane 3.

Figure 16 shows the instantaneous surface pressure in Plane 4, which is near the fan backplate. Consistent pressure distributions are found for all blades. There are no unevenly distributed pressure zones or unsteady fluctuations near the backplate. This phenomenon suggests that the inlet-gap turbulence has negligible interaction with the blades at this location.

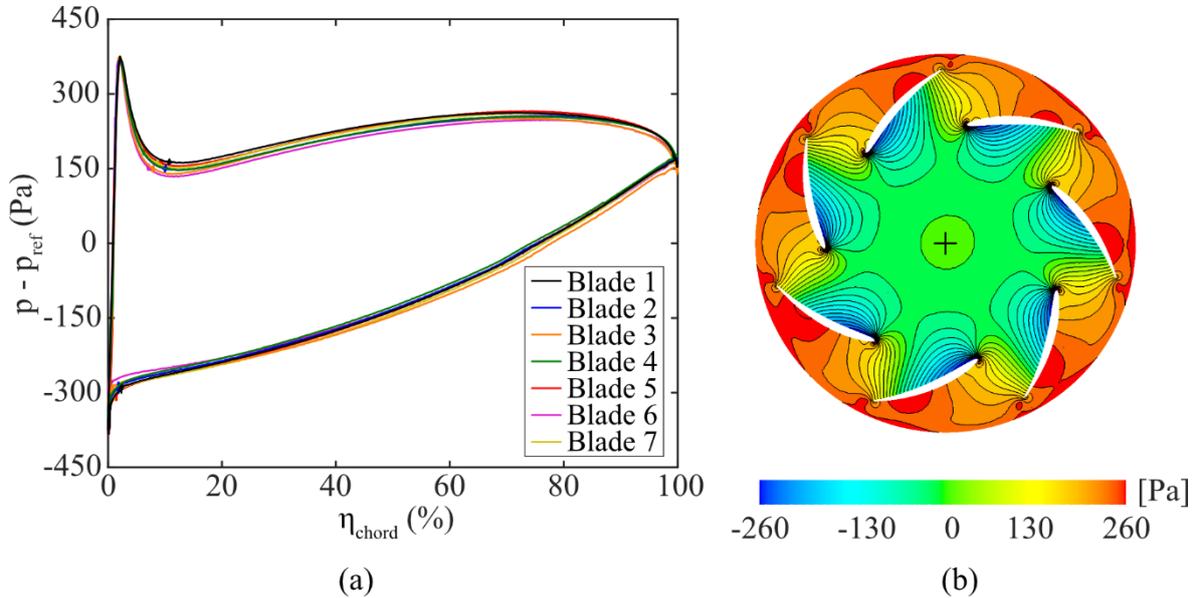

Fig. 16. a) The instantaneous surface pressure a) along the intersection lines of the blades and Plane 4, which is near the fan backplate, and b) in Plane 4.

### 5.3 Turbulence-induced tonal noise

The PSD predicted at the microphones, M1 near the inlet and M2 near the outlet, is compared with the experiments in Fig. 17. The predicted tonal frequencies, $BPF_0$ and $BPF_1$, agree well with the experimental results in all simulations (Case 2 and Case 4). Moreover, a new tonal frequency of 273 Hz is only captured with the refined mesh, but it is not shown using the coarse mesh. The only difference between the refined and coarse meshes is the mesh resolution near the fan inlet gap and shroud. As demonstrated in the aforementioned analysis of flow structures, the refined mesh resolution is essential for resolving the gap turbulence. In other words, the gap turbulence is artificially turned off by coarsening the mesh quality near the inlet gap. Therefore, this specific tone at 273 Hz is generated by the inlet-gap turbulence. This frequency is close to the difference between $BPF_0$ (326.7 Hz) and the fan rotation frequency (46 Hz). It means that the gap turbulence has inertial momentum in the fan rotation direction and interacts with the blades. Moreover, the characteristic rotation velocity of the gap turbulence is nearly equal to the fan rotation speed.

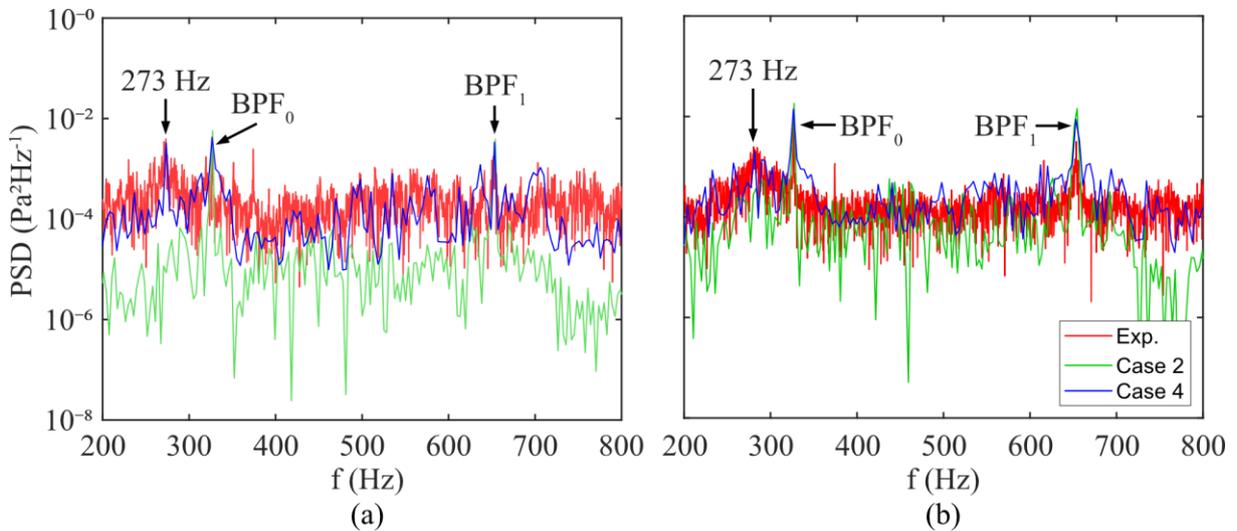

Fig. 17. PSD of the sound pressure at the microphones a) M1 near the inlet and b) M2 near the outlet. The tonal frequencies $BPF_0 = 326.7$ Hz and $BPF_1 = 653.4$ Hz.

The PSD of the sound pressure from the fan surface components (the fan shroud, blades, and backplate) at the microphones, M1 and M2, is displayed in Fig. 18. The tonal frequency $BPF_0$ is observed for all surface components, whereas 273 Hz associated with the gap turbulence exists for the blades and shroud. The tone from the blades is larger than that from the shroud. This effect is understandable in terms of the turbulence kinetic energy and the turbulence impingement direction. The turbulence evolves to be potent at the downstream location near the blade leading edge. The turbulence sweeping direction is parallel to the shroud surface but normal to the blade leading edge. The normal sweeping impingement accounts for more significant pressure and density fluctuations. It is interesting to note that the fan backplate does not present a tone at 273 Hz. The reason is that the gap turbulence disappears near this surface, as indicated in Figs 10 and 14.

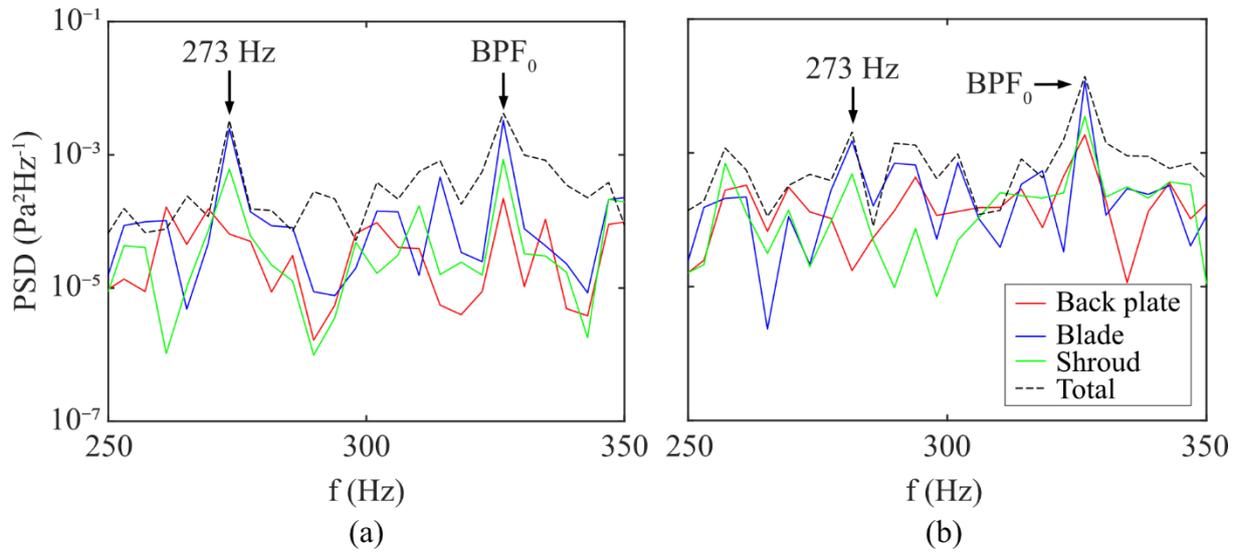

Fig. 18. PSD of the noise generated by the individual fan surface components at a) M1 and b) M2.

### 5.4 Spectral analysis of tonal noise sources

Based on the band filtered PSD of surface pressure and density fluctuations, the locations and magnitudes of dominant tonal noise sources are evaluated. The results at the tonal frequency 273 Hz associated with the gap turbulence are illustrated in Fig. 19. In Case 2 where the gap turbulence is turned off, powerful pressure and density PSD appear only on the top side of the blade trailing edges. In addition to this location, however, Case 4 with the gap turbulence is noticed with plenty of powerful PSD on the inlet gap, the top side of the blade leading edge, and the fan shroud. In particular, large magnitudes are commonly located at the blade leading edges close to the shroud. These high energy locations are identical to the places, where the gap turbulence evolves and accounts for the impingement on the walls (see Figs. 7 and 10). Therefore, the gap turbulence dominates the tonal noise generation at 273 Hz.

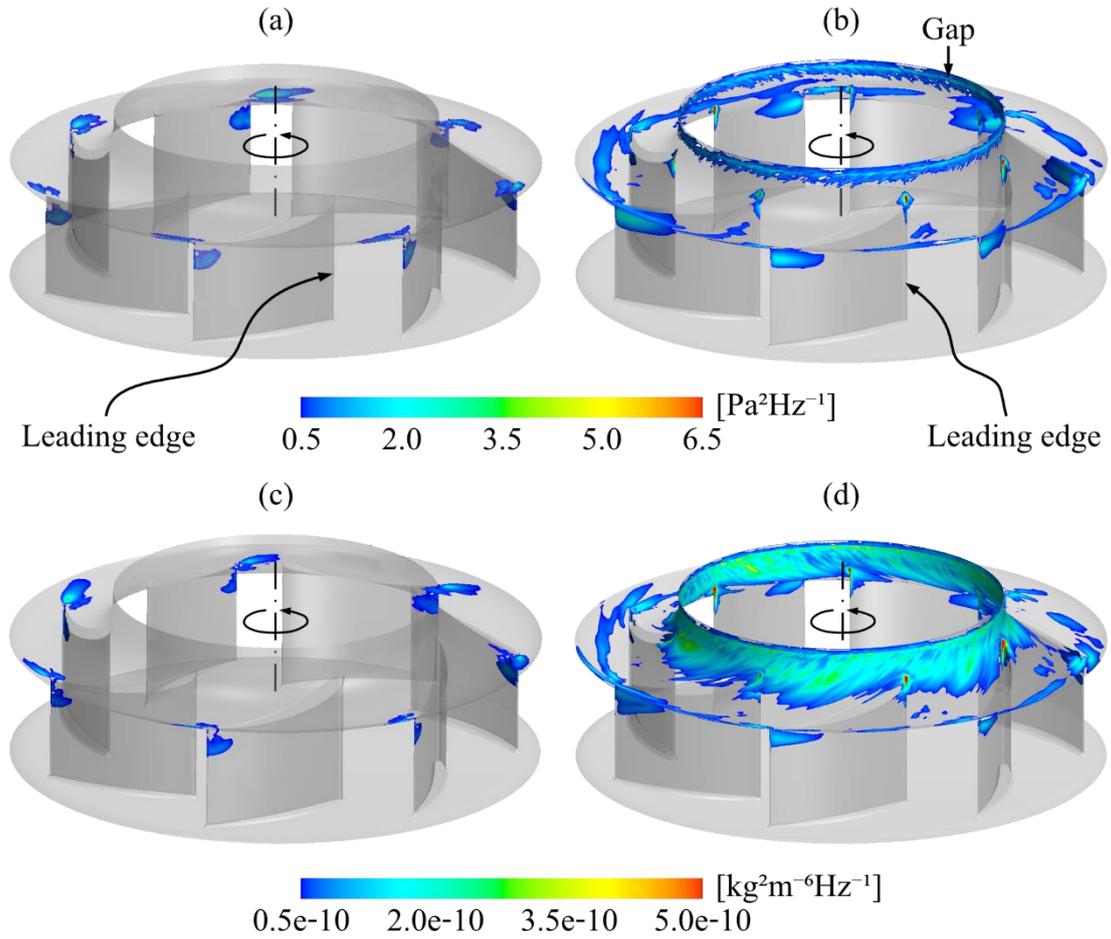

Fig. 19. PSD at 273 Hz for surface pressure fluctuations (top row) and density fluctuations (bottom row). Here a) and c) are for Case 2; and b) and d) are for Case 4. The gap turbulence is artificially turned off in Case 2 and turned on in Case 4. Contours below the minimum values of the color bars are clipped.

## 6 Conclusions

The tonal noise generated from a generic voluteless centrifugal fan for HVAC systems is studied at a specific off-design operation point. The study is motivated by the experimental finding that an increase of the tonal noise is noticed at this operation point, as compared with others. The flow is simulated by a hybrid method coupling the IDDES with Formulation 1A of the FW-H acoustic analogy. The simulated fan performance and noise agree well with the measurement data. Regarding the turbulence stemming from the fan inlet gap, it is the first time to explore its noise generation for voluteless centrifugal fans. A simulation case with coarse mesh resolution at the inlet gap is proposed to artificially deactivate the gap turbulence. By comparing this case to the one that resolves the gap turbulence with refined mesh resolution, the effects of the gap turbulence are addressed.

The gap turbulence evolves along with the fan shroud and is swept downstream to impinge on the fan blades. It accounts for uneven and unsteady surface pressure distribution among the blades. Moreover, the most significant impingement is found on the top side of the blade leading edges, which are near the shroud. The surface pressure in this region fluctuates in a wide amplitude range and is rich in high-frequency contents.

Another effect is that the gap turbulence fades from the blade spanwise top side to bottom side and, finally, disappears near the fan backplate. As the distance to the blade top side increases, surface pressure fluctuations decay rapidly, and similar pressure distribution on different blades is found. However, a periodic fluctuation at the fan rotation frequency is clearly observed with the increased distance.

Spectral analysis is performed for the surface pressure and density fluctuations, and the sound pressure at the microphones near the fan inlet and outlet. The gap turbulence impingement on the shroud and blades contributes to noticeable tonal noise at 273 Hz. Given the fact that this tone disappears when the gap turbulence is artificially turned off, it is concluded that the tone is completely generated from the gap turbulence. The frequency of 273 Hz is close to the difference of $BPF_0$ and the fan rotation frequency (46 Hz). In contrast to conventional blade vortex interaction in which the ambient medium of vortices is quiescent, a specific effect for the present voluteless fan is that the rotating shroud wall introduces rotational momentum into the gap turbulent vorticies due to the wall friction, as the vortices exist near the shroud wall. This is the reason why the tone from the gap turbulence is 273 Hz.

## Acknowledgments

Swegon Operation finances the present work. The simulations were performed on resources provided by the Swedish National Infrastructure for Computing (SNIC) at C3SE.

## Credit authorship contribution statement

**Martin Ottersten:** Conceptualization, Methodology, Software, Analysis, Investigation, Writing – original draft, Writing – review and editing, Visualization. **Hua-Dong Yao:** Methodology, Analysis, Validation, Writing – review and editing, Resources, Supervision. **Lars Davidson:** Writing – review and editing, Resources, Supervision.